\documentclass[aps,prl,floats,twocolumn,showpacs,superscriptaddress]{revtex4}

\usepackage{graphicx,epsfig}
\usepackage{times}
\usepackage{graphics,dcolumn,bm,fleqn,epic,eepic,float}
\usepackage{amssymb,amsmath,multirow,rotate,color}
\def\Erdos{Erd\"os}
\begin{document}

\title{Paths to Synchronization on Complex Networks}

\author{Jes{\'u}s G{\'o}mez-Garde\~{n}es}

\affiliation{Institute for Biocomputation and Physics of Complex
Systems (BIFI), University of Zaragoza, Zaragoza 50009, Spain}

\affiliation{Departamento de F\'{\i}sica de la Materia Condensada,
University of Zaragoza, Zaragoza E-50009, Spain}

\author{Yamir Moreno}

\affiliation{Institute for Biocomputation and Physics of Complex
Systems (BIFI), University of Zaragoza, Zaragoza 50009, Spain}

\author{Alex Arenas}

\affiliation{Departament d'Enginyeria Inform{\`a}tica i
  Matem{\`a}tiques, Universitat Rovira i Virgili, 43007 Tarragona,
  Spain}

\date{\today}

\begin{abstract}

The understanding of emergent collective phenomena in natural and
social systems has driven the interest of scientists from different
disciplines during decades. Among these phenomena, the synchronization
of a set of interacting individuals or units has been intensively
studied because of its ubiquity in the natural world. In this paper,
we show how for fixed coupling strengths local patterns of
synchronization emerge differently in homogeneous and heterogeneous
complex networks, driving the process towards a certain global
synchronization degree following different paths. The dependence of
the dynamics on the coupling strength and on the topology is
unveiled. This study provides a new perspective and tools to
understand this emerging phenomena.

\end{abstract}


\maketitle

In 1998 Watts and Strogatz in an effort to understand the
synchronization of cricket chirps, which show a high degree of
coordination over long distances as though the insects where
``invisibly" connected, end up with a seminal paper about the
small-world effect \cite{watts} that was the seed of the modern theory
of complex networks \cite{strogatz,newmanrev,yamirrep}. Many
natural and man-made networks have been, since then, successfully
described within this framework. Nevertheless, the understanding of the
synchronization dynamics in complex networks remains a challenge.

The synchronization of non-identical interacting units occupies a
privileged position among emergent collective phenomena because of its
various applications in Neuroscience, Ecology, Earth Science, among
others \cite{winfree,strogatzsync,zanette}. One of the most successful
attempts to understand it is due to
Kuramoto \cite{kuramoto75,kurabook}, who analyzed a model of phase
oscillators coupled through the sine of their phase differences. The
Kuramoto model (KM) consists of a population of $N$ coupled phase
oscillators where the phase of the $i$-th unit, denoted by
$\theta_i(t)$, evolves in time according to
\begin{equation}
\frac{d\theta_i}{dt}=\omega_i + \sum_{j}
\Lambda_{ij}A_{ij}\sin(\theta_j-\theta_i) \hspace{0.5cm} i=1,...,N
 \label{ks}
\end{equation}
\noindent where $\omega_i$ stands for its natural frequency,
$\Lambda_{ij}$ is the coupling strength between units and $A_{ij}$ is
the connectivity matrix ($A_{ij}=1$ if $i$ is linked to $j$ and $0$
otherwise). The original model studied by Kuramoto assumed mean-field
interactions with $A_{ij}=1, \forall i\neq j$ (all-to-all) and
$\Lambda_{ij}={\cal K}/N, \forall i,j$. The model can be solved in
terms of an order parameter $r$ that measures the extent of
synchronization in a system of $N$ oscillators as:
\begin{equation}
re^{i\Psi}=\frac{1}{N}\sum_{j=1}^{N} e^{i\theta_j}
 \label{r_kura}
\end{equation}
\noindent where $\Psi$ represents an average phase of the system. The
parameter $0\le r \le 1$ displays a second order phase transition in
the coupling strength, being $r=0$ the value of the incoherent
solution, and $r=1$ the value for total synchronization.

The synchronization problem has been solved in some other cases,
mainly those where a mean-field approach is also valid
\cite{conradrev}. Unfortunately, the mean-field approach requires of
several constraints that are not usually fulfilled in real
systems. Natural, social and technological systems show intricate
patterns of connectivity between their units that are, nowadays,
described as complex networks \cite{newmanrev,yamirrep}. The problem
of synchronization in complex networks inherits the technical
difficulties of the non mean-field approaches and incorporates new
questions to be considered: What are the new pertinent parameters to
deal with synchronization?  and, What is the role of the topology in
the synchronization process? Several works have partially
addressed these issues by studying the stability of the synchronized
state \cite{barahona,motter1,hong,motter2,prl_stefano,munozprl,zhou}
using the Master Stability Function (MSF) formalism
\cite{pecora}. However, the onset of synchronization, which posses
more theoretical and phenomenological challenges is much less
explored, and only a few works have dealt with the study of the whole
synchronization dynamics in specific scenarios
\cite{yamir,kahng,restrepo,arenas,zk}.

The main goal of this Letter is to study the synchronizability of
complex networks as a function of the coupling strengths. To do this,
first, we propose and discuss a new measure of synchronization for the
KM in complex networks. Second, we scrutinize and compare the
synchronization patterns in \Erdos-Renyi (ER) and scale-free (SF)
networks and show that even in the incoherent solution, $r=0$, the
system self-organizes towards synchronization following different
paths. Our study reveals that the synchronizability of these networks
does depend on the coupling between units, and hence, that general
statements about their synchronizability are eventually
misleading. Moreover, we show that even in the incoherent solution,
$r=0$, the system is self-organizing towards synchronization.

Let us start by mapping the KM model in finite complex networks as
\begin{equation}
\frac{d\theta_i}{dt}=\omega_i + \lambda \sum_{j=1}^N
A_{ij}\sin(\theta_j-\theta_i) \hspace{0.5cm} i=1,...,N
 \label{kscn}
\end{equation}
\noindent where $\lambda$ is a constant. 

We study the dynamics of Eq.(\ref{kscn}) in ER and SF networks,
preserving the total number of links, $N_{l}$ and nodes, $N$ for a
proper comparison \cite{note}. We concentrate in two aspects: global
and local synchronization. First, we follow the evolution of the order
parameter $r$, as $\lambda$ increases, to capture the global coherence
of the synchronization in the networks. Secondly, we propose and
follow the same evolution for a new parameter, $r_{link}$. This
parameter measures the local construction of the synchronization
patterns and allows for the exploration of how global synchronization
is achieved. We define
\begin{equation}
r_{link}=\frac{1}{2N_l}\sum_{i}\sum_{j\in \Gamma_i}\left
  |\lim_{\Delta t\rightarrow\infty}\frac{1}{\Delta t}\int_{t_r}^{t_r+\Delta t}e^{i(\theta_i(t)
  -\theta_j(t))}dt\right |,
 \label{r_link}
\end{equation}
\noindent that represents the fraction of all possible links that are
synchronized in the network (averaged over a large enough time
interval $\Delta t$, after the system relaxes at some large time
$t_r$), being $\Gamma_i$ the set of neighbors of node $i$.

\begin{figure}[t]
\begin{center}
\epsfig{file=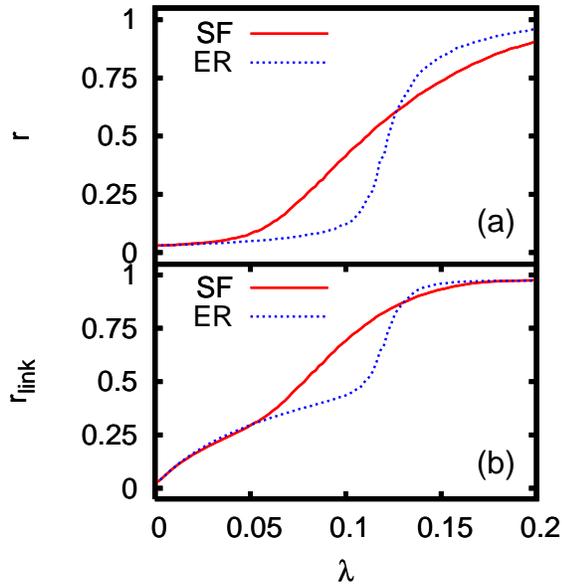,width=3.0in,angle=0,clip=1}
\end{center} 
\caption{(color online) Evolution of {\bf a}, the KM order parameter
defined in Eq.\ (\ref{r_kura}), and {\bf b} the fraction of
synchronized links $r_{link}$, Eq.\ (\ref{r_link}), as a function of
$\lambda$. The curves separate when the incoherent solution for SF
networks destabilizes. The figure clearly illustrates that the
synchronizability of the networks does depend on the value of the
coupling strength. Both plots are represented for \Erdos-Renyi (ER)
and scale-free (SF) networks as indicated. The size of the networks is
$N=1,000$ and their average degree is $\langle k \rangle=6$. The
exponent of the SF network is $\gamma=-3$.}
\label{R}
\end{figure}

\begin{figure}
\begin{center}
\epsfig{file=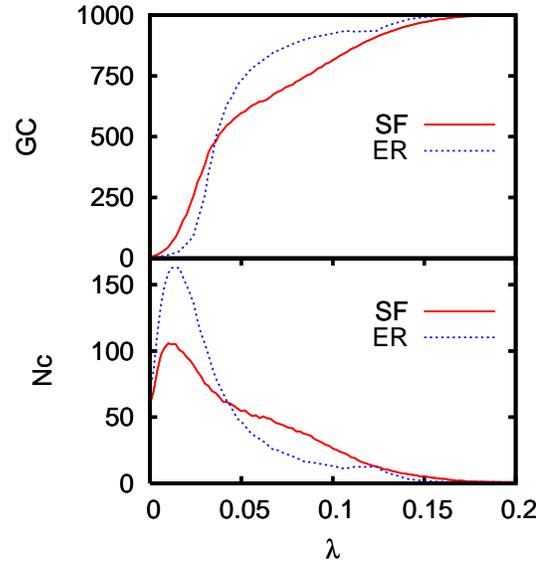,width=3.0in,angle=-90,clip=1}
\end{center} 
  \caption{(color online) Size of largest synchronized connected
  component ($GC$) and number of synchronized connected components
  ($Nc$), as a function of $\lambda$ for the different topologies
  considered. Despite $r$ being vanishing and hence no global
  synchronization is yet attained, a significant number of clusters
  show up. This indicates that for any $\lambda>0$ the system
  self-organizes towards macroscopic synchronization. The network
  parameters are as in Fig.\ \ref{R}.}
\label{pattern}
\end{figure}

We solved Eq.(\ref{kscn}) using a $4^{th}$ order Runge-Kutta method
for different values of $\lambda$, with a uniform distribution of
natural frequencies $g(\omega)$ in the interval $[-\pi,\pi]$ up to
achieving the stationary state. The networks are built following a
model \cite{jesus} that generates a one parameter family of complex
networks. This parameter, $\alpha \in [0, 1]$, measures the degree of
heterogeneity of the final networks. A network of size $N$ is
generated starting from a fully connected core of $m_{0}$ nodes and a
set ${\cal U}(0)$ of $N-m_{0}$ unconnected nodes. At each time step, a
new node (not selected before) is chosen from $ {\cal U}(0)$ and
linked to $m$ other nodes. Each of the $m$ edges is linked with
probability $\alpha$ to a randomly chosen node (avoiding multiple and
self-connections) from the whole set of $N-1$ remaining nodes and with
probability $(1-\alpha)$ following a linear preferential attachment
strategy \cite{doro}. Repeating these steps $(N-m_{0})$ times,
networks interpolating between the limiting cases of ER ($\alpha=1$)
and SF ($\alpha=0$) topologies are generated \cite{note1}.

In Fig.\ \ref{R} we represent the evolution of both order parameters,
$r$ and $r_{link}$, as a function of the coupling strength $\lambda$.
The global coherence of the synchronized state, represented by $r$,
shows that the onset of synchronization first occurs for SF
networks. A detailed finite size scaling analysis performed for both
topologies shows that the critical value of the effective coupling,
$\lambda_c$, corresponds in SF networks to $\lambda_c^{SF} = 0.05(1)$,
and in ER networks to $\lambda_c^{ER} = 0.122(2)$, accordingly with
Fig.\ \ref{R}. If $\lambda$ is further increased, there is a value at
which $r$ for the ER crosses over the SF curve. From this value up in
$\lambda$, the ER network remains slightly more synchronized than the
SF network.

The behavior of $r_{link}$ shows a change in synchronizability
between ER and SF and provides additional information. Interestingly,
the nonzero values of $r_{link}$ for $\lambda\leq\lambda_c$ indicate
the existence of some local synchronization patterns even in the
regime of global incoherence ($r \approx 0$). Right at the onset of
synchronization for the SF network, its $r_{link}$ value deviates from
that of the ER. While the synchronization patterns continue to grow
for the ER network at the same rate, the formation of locally
synchronized structures occurs at a faster rate in the SF
network. Finally, when the incoherent solution in the ER network
destabilizes, the growing in its synchronization pattern increases
drastically up to values of $r_{link}$ comparable to those obtained in
SF networks and even higher.

The above results show that statements about synchronizability are
dependent on the coupling strength value. Additionally, the previous
discussion suggests that synchronization is attained following two
different paths that depend on the underlying topology. We have
studied the characteristics of the synchronization patterns along the
evolution of $r_{link}$.  Synchronization patterns are formed by pairs
of oscillators, physically connected, whose phase difference in the
stationary state tends to zero. Note that the contribution into
Eq.(\ref{r_link}) of every pair of connected oscillators can be
written in terms of a matrix ${\cal D}_{ij} =A_{ij}\left |\lim_{\Delta
t\rightarrow\infty}\frac{1}{\Delta t}\int_{t_r}^{t_r+\Delta
t}e^{i(\theta_i(t) -\theta_j(t))}dt\right |$. This matrix is filtered
using a threshold $T$ such that the fraction of synchronized pairs
equals $r_{link}$. In this way, if ${\cal D}_{ij}>T$ oscillators $i$
and $j$ are considered synchronized and the synchronized patterns are
extracted.

\begin{figure}
\begin{center}
\epsfig{file=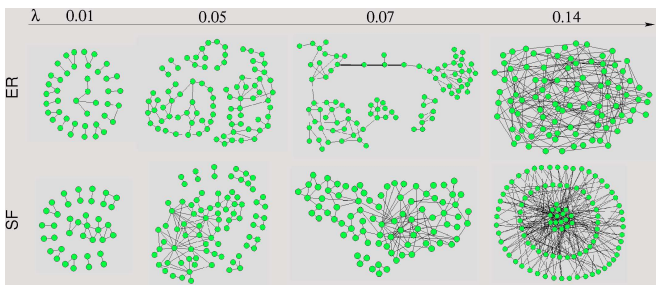,width=\columnwidth}
\end{center}
  \caption{(color online) Synchronized clusters for several values of
  $\lambda$ for the two different topologies studied (ER and
  SF). These networks are made up of 100 nodes, in order to have a
  sizeable picture of the system. The evolution of local
  synchronization patterns is always agglomerative, however, it
  follows two different routes. In the ER case, the growth of the GC
  proceeds by aggregation of small clusters of synchronized nodes,
  while for the SF network the central core groups the smaller
  clusters around it.}
\label{nets}
\end{figure}

In Fig.\ \ref{pattern} we represent the number of synchronized clusters and the
size of the largest one (GC) as a function of $\lambda$. The local
information extracted from it is unveiling an astonishing and novel
feature of the synchronization process that can not be derived from
Fig.\ \ref{R}, and that in some sense is counterintuitive. The emergence of
clusters of synchronized pairs of oscillators (links) in the networks
shows that for values of $\lambda \le \lambda_c^{SF}$, i.e., still in
the incoherent solution $r=0$, both kind of networks have developed a
largest cluster of synchronized pairs of oscillators involving $50\%$
of the nodes of the network, and an equal number of smaller
synchronization clusters. From this point on, in the SF network the GC
grows and the number of smaller clusters goes down, whereas for the ER
network the growth exploits. These results indicate that although SF
networks present more coherence in terms of $r$ and $r_{link}$, the
microscopic evolution of the synchronization patterns is faster in ER
networks, being these networks far more locally synchronizable than
the heterogeneous ones.

\begin{figure}[t]
\epsfig{file=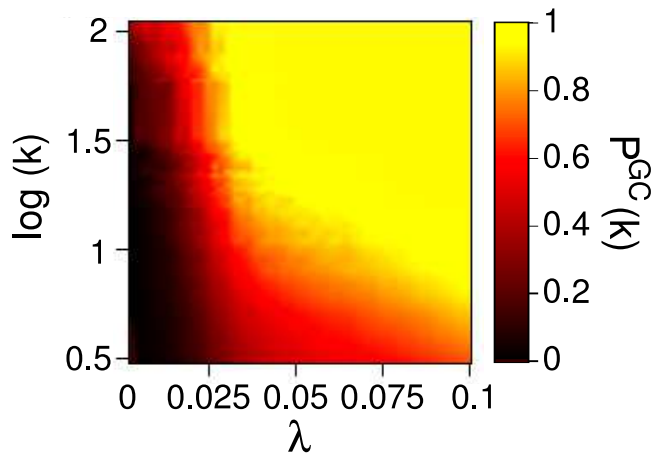,width=\columnwidth,angle=0,clip=1}
  \caption{(color online) The plot shows the correlation between the likelihood
  that a node belongs to the GC of pairs of synchronized oscillators
  and its degree $k$ as a function of the coupling strength
  $\lambda$. This probability ,$P^{GC}(k)$, is color-coded as
  indicated in the right panel. The figure convincingly demonstrates
  that highly connected nodes are those that recruit poorly connected
  nodes as the GC grows. Network parameters are those used in Fig.\ \ref{R}}
\label{kint}
\end{figure}

The observed differences in the behavior at a local scale are rooted
in the growth of the GC. It turns out that for the ER networks, many
different clusters of synchronized pairs of oscillators merge together
to form a GC when the effective coupling is increased. The coalescence
of many small clusters leads to a giant component of synchronized
pairs that is almost the size of the system once the incoherent state
destabilizes. This is not anymore the case for SF networks, where
oscillators are incorporated to the GC practically one-by-one (forming
new pairs) in terms of $\lambda$ (or $r_{link}$), but starting from a
core made up of half the nodes of the network. This picture is
confirmed in Fig.\ \ref{nets}, where we have represented the evolution
of local synchronization patterns in ER and SF networks for several
values of $\lambda$. The ultimate reason behind these two different
routes to complete synchronization is the heterogeneous character of
the SF network and the role played by the hubs. In Fig.\ \ref{kint},
we have plotted the probability that a node with degree $k$ belongs to
the GC as a function of its degree $k$ and the coupling $\lambda$ for
the SF network. This probability is an increasing function of $k$ for
every $\lambda$, hence the more connected a node is, the more likely
it takes part in the cluster of synchronized links. Recently
\cite{zk}, Zhou and Kurths have reported the study of hierarchical
organization in complex networks, using the MSF and a mean-field
approach in the weak coupling limit. Our results thus substantiate and
generalize those about the role of hubs in the synchronization process
presented in \cite{zk}.

In summary, we have shown that synchronizability of complex networks
is dependent on the effective coupling $\lambda$ among
oscillators. For small values of $\lambda$, SF networks outperform ER
topologies, but the tendency is reverted for intermediate to large
values of the coupling. On the other hand, the detailed analysis of
evolution of patterns of synchronization showed that there are two
radically different mechanisms to attain synchronization. In the
presence of hubs, a giant component of synchronized pairs of
oscillators forms and grows by recruiting nodes linked to them. On the
contrary, in homogeneous structures, many small clusters first appear
and then group together through a sharp merging process. These results
are as far reaching as the ones obtained for percolation and epidemic
spreading on top of homogeneous or heterogeneous graphs, where the
radical differences of the system's dynamics are rooted in the
topology of the underlying networks, demonstrating that the same
behavior may hold for nonlinear dynamical systems coupled to complex
structures. However, at variance with percolation processes, here the
synchronization patterns could be directly related to the growth and
evolution of the network. Therefore, we have naturally incorporated a
dynamics relevant for the emergence of cooperative behavior, showing
that the same organizing principle may drive network evolution, i.e.,
if synchronization is a relevant issue, natural networks (whether they
are homogeneous or heterogeneous in degree) can indeed be efficient by
adaptively selecting their coupling strengths. Our study then opens
new paths to clarify how synchronization is attained in complex
topologies and provides new tools to analyze this ubiquitous
phenomenon.

\begin{acknowledgments}
  We thank J.A. Acebr\'on, A. D\'{\i}az-Guilera,
  C.J. P\'{e}rez-Vicente and V. Latora for helpful
  comments. J.G.G. and Y.M. are supported by MEC through a FPU grant
  and the Ram\'{o}n y Cajal Program, respectively. This work has been
  partially supported by the Spanish DGICYT Projects
  FIS2004-05073-C04-01 and FIS2005-00337.
\end{acknowledgments}


\begin{thebibliography}{91}
\bibitem{note} Note that for a proper comparison of the
synchronizability of different complex networks, the global and local
measures of coherence should be represented according to their
respective time scales. Therefore, given two complex networks A and B
with $k_{max}=k_A$ and $k_{max}=k_B$ respectively, the comparison
between observables must be done for the same effective coupling
$\frac{\lambda_A}{k_A}=\frac{\lambda_B}{k_B}=\lambda$.
\bibitem{note1} The study reported henceforth has also been conducted
for intermediate values of $\alpha$. Starting from $\alpha=0$ and
increasing its value, the networks generated are successively more
homogeneous \cite{jesus} with a heavy tail whose exponent is equal to
($\alpha=0$) or larger than ($\alpha>0$) $\gamma=3$. We advance that
the results for $0 < \alpha < 1$ are consistent with the picture
described in the rest of the paper.
\end{thebibliography}
\end{document}